\documentclass[twocolumn,showpacs,preprintnumbers,prl,amsmath,amssymb]{revtex4}

\usepackage{graphicx}
\usepackage{dcolumn}
\usepackage{bm}

\begin{document}

\title{Magnetic and lattice coupling in single-crystal SrFe$_2$As$_2$: A neutron scattering study}

\author{Haifeng Li, Wei Tian, Jerel L. Zarestky, Andreas Kreyssig, Ni Ni, Sergey L. Bud$'$ko, Paul C. Canfield, Alan I. Goldman, Robert J. McQueeney and David Vaknin}

\affiliation{Ames Laboratory and Department of Physics and Astronomy, Iowa State University, Ames, Iowa 50011, USA}


\date{\today}

\begin{abstract}

A detailed elastic neutron scattering study of the structural and magnetic phase transitions in single-crystal SrFe$_2$As$_2$ reveals that the orthorhombic (O)-tetragonal (T) and the antiferromagnetic transitions coincide at    $T_\texttt{O}$ = $T_\texttt{N}$ = (201.5 $\pm$ 0.25) K. The observation of coexisting O-T phases over a finite temperature range at the transition and the sudden onset of the O distortion provide strong evidences that the structural transition is first order. The simultaneous appearance and disappearance within 0.5 K upon cooling and within 0.25 K upon warming, respectively, indicate that the magnetic and structural transitions are intimately coupled. We find that the hysteresis in the transition temperature extends over a 1-2 K range. Based on the observation of a remnant orthorhombic phase at temperatures higher than \emph{T}$_\texttt{O}$, we suggest that the T-O transition may be an order-disorder transition.

\end{abstract}

\maketitle

\section{INTRODUCTION}

The newly discovered iron-based superconductors \emph{Ln}FeAs(O$_{1-x}$F$_x$) (\emph{Ln} = Lanthanides) \cite{Kamihara2008,Chen2008}, and oxygen-free doped \emph{A}Fe$_2$As$_2$ (\emph{A} = Ca, Sr, Ba, Eu, K, Cs, Li) \cite{Pfisterer1980, Yan2008, Ni2008, Sasmal2008}, LiFeAs \cite{Wang2008}, FeSe \cite{Mizuguchi2008} and SrFeAsF \cite{Wu2008} have stimulated a great deal of activity on superconductors derived from antiferromagnetic (AFM) parent compounds. This novel class of materials, besides existing cuprate-based high-$T_\texttt{c}$ superconductors, provides yet another system for exploring the interplay between superconductivity and antiferromagnetism. While for cuprates, the Cu-O planes are crucial to the superconducting (SC) behavior, in so-called $\texttt{"}$122$\texttt{"}$ iron-based superconductors, the FeAs layers play a similar role. Suitable hole- or electron-doping as well as applied pressure can suppress both structural and magnetic transitions resulting in superconductivity with $T_\texttt{c}$ up to 55 K \cite{Rotter2008, Sefat2008, Torikachvili2008}. Due to the weak electron-phonon interactions and the emergence of superconductivity with a disappearance of the AFM transition, spin fluctuations have been proposed to play a key role in establishing the SC state in these iron-based supercondcutors \cite{Haule2008, Singh2008}. However, in these compounds the AFM state is strongly coupled to a lattice distortion and it is, therefore, important to unravel the interactions between lattice and spin degrees of freedom.

\begin{figure} [htl]
\centering
\includegraphics[width = 0.23\textwidth] {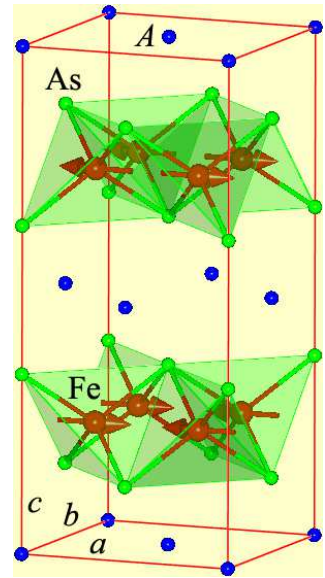}
\caption{\label{structure}(color online).
Crystal (\emph{Fmmm}) and AFM structure of \emph{A}Fe$_2$As$_2$ (\emph{A} = Ca, Sr, and Ba) below \emph{T}$_\texttt{O}$.
}
\label{UnitStructure1}
\end{figure}

The parent compounds of \emph{A}Fe$_2$As$_2$ adopt a tetragonal ThCr$_2$Si$_2$-type structure at room temperature and undergo a tetragonal- (T) to-orthorhombic (O) phase transition upon cooling. This transition is accompanied by an AFM ordering with magnetic moments aligned parallel to the crystallographic \emph{a} axis with a propagation vector along the [101] (O notation) direction (Fig.\ \ref{structure}). The majority of reports \cite{Ni2008, Goldman2008, Krellner2008, Zhao2008} characterize the O-T transition as a first-order (FO) structural transition with small hysteresis, while some \cite{Zhao2008, Tegel2008} have argued that the accompanying AFM transition seems to be continuous, or more difficult to characterize as first- or second-order. However, neutron-diffraction measurements of single-crystal CaFe$_2$As$_2$ clearly show that the AFM transition can be classified as a first-order transition with a $\sim$1 K hysteresis \cite{Goldman2008}.

The synthesis, structure, and magnetic susceptibility measurements of SrFe$_2$As$_2$ were reported by Pfisterer \cite{Pfisterer1980}in 1980, where an anomaly in the susceptibility around 200 K was associated with an AFM transition. This anomaly was further characterized as a FO transition at $T_\texttt{O} = T_\texttt{N} = 205$ K in polycrystalline SrFe$_2$As$_2$ samples by resistivity and specific-heat measurements \cite{Krellner2008}. Single-crystal studies of SrFe$_2$As$_2$ showed a structural transition from a high-temperature T (\emph{I}4/\emph{mmm}) phase to a low-temperature O (\emph{Fmmm}) one at $T_\texttt{O}$, simultaneously accompanied by the AFM transition \cite{Yan2008, Zhao2008}. The neutron-scattering study of a single-crystal SrFe$_2$As$_2$ (\emph{T}$_\texttt{O}$ = 220 K) \cite{Zhao2008} determined that the AFM spin direction is parallel to the crystalline \emph{a} axis in \emph{Fmmm} symmetry and that the structural transition is first order, while the magnetic transition appears to be continuous. However, the hysteresis of both transitions was not reported. By contrast, a powder x-ray diffraction study reported that the \emph{A}Fe$_2$As$_2$ compounds (\emph{A} = Ba, Sr, Ca) undergo second-order displacive structural transitions \cite{Tegel2008}.

X-ray diffraction and resistivity measurements under pressure show that \emph{T}$_\texttt{O}$ shifts to lower temperatures and the transition is practically suppressed at a critical pressure range of 4-5 GPa. Above 2.5 GPa, a significant decrease in resistivity below $\sim$40 K indicates the appearance of superconductivity. 40\% Co-doping on the iron site in SrFe$_2$As$_2$ results in the coexistence of superconductivity ($T_\texttt{c}$ = 19.5 K) and an AFM state ($T_\texttt{N}$ = 120 K) \cite{Kim2008}, while 40 \%-50 \% substitution of the Sr site by K and Cs brings $T_\texttt{c}$ to 37 K \cite{Sasmal2008}. As mentioned above, the reported $T_\texttt{N}$ of SrFe$_2$As$_2$ differs among studies, e.g., 198 K (single crystal) \cite{Yan2008}, 200 K (single crystal) \cite{Chen2008b}, 205 K (polycrystal) \cite{Kaneko2008}, and 220 K (single crystal) \cite{Zhao2008}. This is probably due to the subtle changes in FeAs layers resulting from different synthesizing processes, e.g., the Sn incorporation.

In order to clarify the link between magnetism and structure, both transitions have to be monitored simultaneously in one experiment. Neutron scattering is an ideal technique for this kind of measurement. Herein, we report a systematic elastic neutron scattering study on a single-crystal SrFe$_2$As$_2$, focusing on the details of both magnetic and structural transitions especially close to the transition temperature.

\section{EXPERIMENTAL}

\begin{figure} [htl]
\centering
\includegraphics[width = 0.45\textwidth] {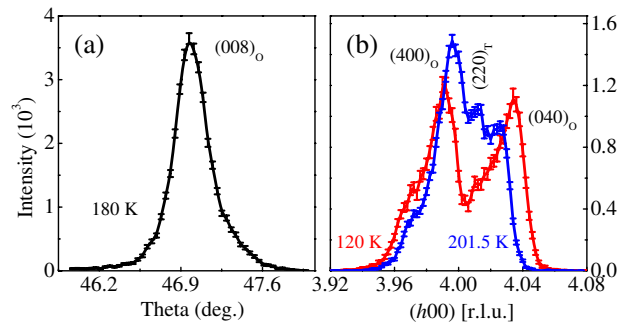}
\caption{\label{rocking}(color online).
(a) Rocking curve of (008)$_\texttt{O}$ indicating the mosaic spread in our crystal. (b) Coexistence of O and T structures at 201.5 K and splitting of (040)$_\texttt{O}$/(400)$_\texttt{O}$ due to twins in \emph{Fmmm} symmetry at 120 K upon warming. O and T stand for orthorhombic and tetragonal, respectively.
}
\label{Rocking1}
\end{figure}

High-quality SrFe$_2$As$_2$ single crystals were synthesized by the FeAs flux growth technique \cite{Ni2008-1}. The crystallinity and purity were characterized by Laue back-scattering, x-ray powder diffraction, magnetization and resistivity measurements. The mosaic of the investigated single crystal is 0.29(1)$^\texttt{o}$ full width at half maxima for the (008)$_\texttt{O}$ reflection [Fig.\ \ref{rocking}a] and the lattice parameters were determined to be \emph{a} = 0.5530(1) nm, \emph{b} = 0.5465(1) nm, and \emph{c} = 1.2213(1) nm in \emph{Fmmm} symmetry at 20 K in this study. The elastic neutron scattering measurements were carried out on the HB-1A fixed-incident-energy (14.6 meV) triple-axis spectrometer using a double pyrolitic graphite (PG) monochromator (high flux isotope reactor at the Oak Ridge National Laboratory, USA). Two highly oriented PG filters, one after each monochromator, were used to reduce the $\lambda$/2 contamination. The beam collimation throughout the experiment was kept at 48$'$-20$'$-sample-20$'$-34$'$. A single crystal ($\sim$15 mg) was wrapped in Al foil and sealed in a He-filled Al can which was then loaded on the cold tip of a closed cycle refrigerator with $(H 0 L)_\texttt{O}$ in the scattering plane. We note that the $(H K L)_\texttt{O}$ indices for O symmetry correspond to the T reflections $(h k l)_\texttt{T}$  based on the relations $H = h - k, K = h + k$, and $L = l$.

\begin{figure} [htl]
\centering
\includegraphics[width = 0.45\textwidth] {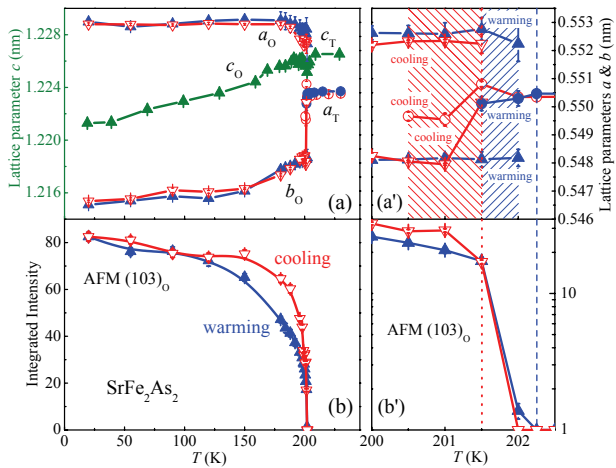}
\caption{\label{LattParam}(color online).
Temperature variations (warming: solid symbols; cooling: void symbols) of (a) the lattice parameters \emph{a}$_\texttt{O}$, \emph{b}$_\texttt{O}$, and \emph{c}$_\texttt{O}$, and \emph{a}$_\texttt{T}$ (circles) and \emph{c}$_\texttt{T}$, (b) the integrated intensity of AFM (103)$_\texttt{O}$ of a single-crystal SrFe$_2$As$_2$. (a$'$) Coexistence region of the O and T phases. (b$'$) Enlarged (b) near the transition temperature. O and T stand for orthorhombic and tetragonal, respectively. The lattice parameter \emph{a}$_\texttt{T}$ in \emph{I}4/\emph{mmm} symmetry was multiplied by $\sqrt[]{2}$ for comparison to the \emph{Fmmm} lattice parameters \emph{a}$_\texttt{O}$ and \emph{b}$_\texttt{O}$.
}
\label{LatticeParameters1}
\end{figure}

\begin{figure} [htl]
\centering
\includegraphics[width = 0.43\textwidth] {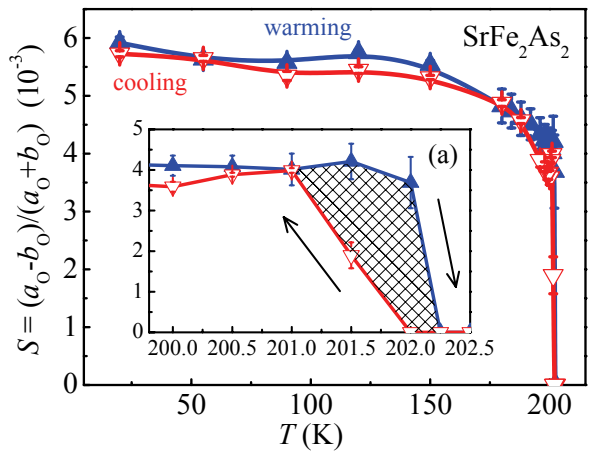}
\caption{\label{Magnetic}(color online).
Temperature variation in the O distortion in the crystalline \emph{a}-\emph{b} plane, namely, \emph{S} = (\emph{a}$_\texttt{O}$-\emph{b}$_\texttt{O}$)/(\emph{a}$_\texttt{O}$+\emph{b}$_\texttt{O}$). Inset (a) is the enlarged figure near the transition.
}
\label{Magnetic1}
\end{figure}

\begin{figure} [htl]
\centering
\includegraphics[width = 0.43\textwidth] {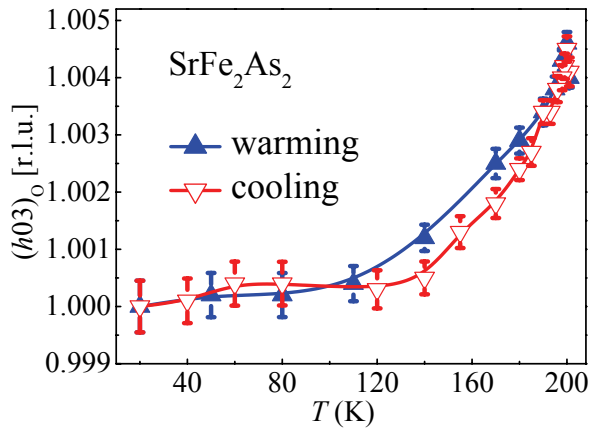}
\caption{\label{shift}(color online).
The change in position of the AFM (103)$_\texttt{O}$ relative to the alignment at 20 K follows the change in the position of the (400)$_\texttt{O}$ related to the temperature dependence of the lattice parameter $a_\texttt{O}$. This temperature dependence is evidence that the direction of the AFM propagation vector is along the long O \emph{a} axis consistent with the observation of the complete absence of (100)$_\texttt{O}$ peak in our experiment. O stands for orthorhombic.
}
\label{shift1}
\end{figure}

\begin{figure} [htl]
\centering
\includegraphics[width = 0.43\textwidth] {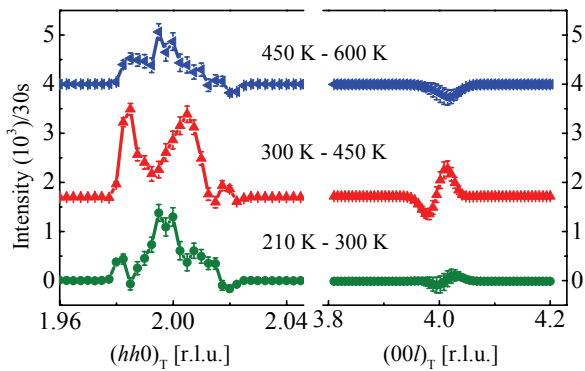}
\caption{\label{highTem}(color online).
Differences of neutron scattered intensities at the $(hh0)_\texttt{T}$ and $(00l)_\texttt{T}$ positions in the tetragonal phase (T stands for tetragonal) with temperature up to 600 K. The residual intensities  at the $(00l)_\texttt{T}$ position suggest that the thermal effect is neglectable. The residual  peak shapes at the $(hh0)_\texttt{T}$ position suggest a remnant orthorhombic phase is present above the O-T transition. The residual intensity at 300 and 450 K was shifted along its positive axis for clarity.
}
\label{highTem1}
\end{figure}

\section{RESULTS AND DISCUSSION}

The splitting of the (220)$_\texttt{T}$ (\emph{I}4/\emph{mmm}) reflection [Fig.\ \ref{rocking}b] into twinned (040)$_\texttt{O}$/(400)$_\texttt{O}$ (\emph{Fmmm}) reflections is a sensitive measure of the T-to-O transition and was used to detect the temperature evolution of the two phases. Lattice parameters \emph{a}$_\texttt{O}$, \emph{b}$_\texttt{O}$, and \emph{c}$_\texttt{O}$ in \emph{Fmmm} symmetry and \emph{a}$_\texttt{T}$ (circles) and \emph{c}$_\texttt{T}$ in \emph{I}4/\emph{mmm} symmetry of the single-crystal SrFe$_2$As$_2$ were obtained from the monitored (400)$_\texttt{O}$/(040)$_\texttt{O}$ or (220)$_\texttt{T}$ and (008)$_{\texttt{O/T}}$ peak positions and are displayed in Fig.\ \ref{LattParam}a. Near the onset of the T-O transition, longitudinal scans through the (220)$_\texttt{T}$ position clearly show the coexistence of O and T phases, while well below the transition (e.g., at 120 K) no coexistence exists [Fig.\ \ref{rocking}b]. These scans were performed for a series of temperatures, both cooling and warming, to study the coexistence behavior. Upon warming, the O-T coexistence region was found to be confined to the 201.5-202 K temperature range, while upon cooling the coexistence region shifts to slightly lower temperatures 200.5-201.5 K [see two shaded regions in Fig.\ \ref{LattParam}a$'$]. Upon warming, the lattice parameters \emph{a}$_\texttt{O}$ and \emph{b}$_\texttt{O}$ abruptly merge into \emph{a}$_\texttt{T}$ at 202.25 K while upon cooling \emph{a}$_\texttt{T}$ forks into \emph{a}$_\texttt{O}$ and \emph{b}$_\texttt{O}$ at 200 K. This is evidence that the structural transition is FO and also indicates that the temperature (T-to-O or O-to-T transition) \emph{T}$_\texttt{O}$ = (201.5 $\pm$ 0.25) K. Below 200 K, \emph{a}$_\texttt{O}$ continuously increases down to $\sim$150 K with cooling and then it saturates within estimated standard deviation (esd), whereas \emph{b}$_\texttt{O}$ decreases smoothly down to 20 K consistent with the trend reported in powder x-ray diffraction studies \cite{Tegel2008}. Except for the coexistence regimes of O and T phases, the lattice parameters vary smoothly within esd upon cooling or warming.

Our neutron diffraction results show that the lattice parameters in \emph{I}4/\emph{mmm} and \emph{Fmmm} symmetries away from $T_\texttt{O}$ evolve without any irregularities as a function of temperature. From the derived lattice parameters of the O phase, we calculated the order parameter in the \emph{a}-\emph{b} plane, namely, O distortion \emph{S} $\equiv$ (\emph{a}$_\texttt{O}$-\emph{b}$_\texttt{O}$)/(\emph{a}$_\texttt{O}$+\emph{b}$_\texttt{O}$), as shown in Fig.\ \ref{Magnetic}. A closer inspection near the transition (inset a) reveals $\sim$1.25 K hysteresis and the distortion is a little larger upon warming than upon cooling. At 201 K upon cooling and 202 K upon warming, \emph{S} has a sharp jump and is at approximately the two-third level of its value at 20 K. Similar weak hysteresis effect has also been reported for CaFe$_2$As$_2$ ($\sim$1 K) \cite{Goldman2008}, while a larger one ($\sim$20 K) was observed in BaFe$_2$As$_2$ \cite{Kofu2009}. It is pointed out that the hysteresis of the O-T transition strongly depends on the temperature history since the intrinsic strain, the external stress or pressure, and defects that may pin the structure locally.

Figure \ref{LattParam}b shows the integrated intensity of the strongest AFM (103)$_\texttt{O}$ reflection as a function of temperature in a warming-cooling cycle after the initial cooling process, with an apparent difference between cooling and warming indicatives of a huge difference between the AFM domain volumes. Figure \ \ref{shift} shows the peak position along the (\emph{h}00) of the AFM (103)$_\texttt{O}$ peak. It follows the same relative temperature-dependent trend that the (400)$_\texttt{O}$ shows, practically following the distortion of the long O \emph{a} axis. This establishes the direction of the in-plane AFM propagation vector unequivocally along the long O \emph{a} axis. The appearance and disappearance of the AFM (103)$_\texttt{O}$ reflection and the O structure occur at exactly the same temperature range (warming: 202-202.25 K; cooling: 201.5-202 K) [Figs.\ \ref{LattParam}a$'$ and b$'$], indicating a strong coupling of the two transitions.

The difference of AFM (103)$_\texttt{O}$ between warming and cooling [Fig. \ref{LattParam}b] is not expected from an ordinarily second-order phase transition. One reason for this difference can be the effect of intrinsic stress/pressure developing on the boundaries of twin domains as a result of the O distortion. Such stress may have a similar effect as pressure, known to lower the AFM transition temperature and even suppresses it \cite{Kumar2008,Yu2009}. Such an effect can lead, for example, to non-monotonic behavior of the order parameter with a monotonic variation in temperature. The field cooling (FC), zero FC and heat treatments also have a strong influence on the magnetic susceptibility in this AFM phase \cite{Saha2008}. A similar behavior has also been reported in a colossal magnetoresistance single-crystal La$_{1-\texttt{x}}$Sr$_\texttt{x}$MnO$_3$ (x $\sim$ 0.125) \cite{Li2009}.

We have conducted a few high-temperature (up to 600 K) measurements, which shows that the residual intensity of $(h h 0)_\texttt{T}$ at 210, 300, and 450 K after subtracting the higher-temperature one at 300, 450, and 600 K, respectively, keeps some remnant splitting (Fig.\ \ref{highTem}) indicative of the existence of O phase above \emph{T}$_\texttt{O}$ even up to 450 K. This may indicate that the O-T transition is, in fact, an order-disorder transition, namely, that the system is O locally even above the \emph{T}$_\texttt{O}$. This is consistent with the observation of strong spin fluctuations above \emph{T}$_\texttt{O}$ up to at least 300 K in the related CaFe$_2$As$_2$ compound \cite{Diallo2009}. This also reports that the T phase can exist below \emph{T}$_\texttt{O}$ in SrFe$_2$As$_2$ \cite{Saha2008}. Defect structures were also observed in other ThCr$_2$Si$_2$-type compounds, e.g., URu$_2$Si$_2$ \cite{Ramirez1991}. In this order-disorder scenario, O and T phases coexist below and above $T_\texttt{O}$. Below \emph{T}$_\texttt{O}$, the O phase is dominating, while above \emph{T}$_\texttt{O}$ the T phase is the majority. The minor phase in both cases may be too weak or the dissimilarity in the structural parameters of both phases is too small to be detected within instrumental resolution \cite{Ma2008, Saha2008}. This also strongly depends on the size of the investigated single crystal for neutron-scattering studies. However, a pseudo-periodic structural modulation at room temperature probably connected with local structural fluctuations and the presence of complex domain structures in the low-temperature O phase were indeed revealed in a TEM study \cite{Ma2008}. The minor T phase below \emph{T}$_\texttt{O}$ may pin the AFM domains, preventing their enlargement and rotation resulting in the continuous change in the observed  AFM domains. The effect of an intrinsic disorder on the AFM domain structure was also suggested in Ref. \cite{Saha2008}. The small amount of T phase below \emph{T}$_\texttt{O}$ may be also the source of the reported phase separation of orthorhombic-striped magnetic clusters and tetragonal SC clusters \cite{Ricci2008}. The small value of the AFM moments (neutron diffraction: \cite{Zhao2008} 0.94 $\mu$$_\texttt{B}$; $\texttt{LDA-SDW}$: \cite{Kasinathan2009} 1.13 $\mu$$_\texttt{B}$; $\mu$SR: \cite{GoKo2008} 0.8 $\mu$$_\texttt{B}$) may indicate a regional spin-frozen state due to the anisotropy of possible clusters. The minor T phase could be the source for the formation of magnetic clusters.

To summarize, employing the neutron-diffraction technique to characterize the nature of the magnetic and structural transitions we found the following major features which we supplement with our assessments: (1) the structural distortion and the AFM ordering coalesce into a single-phase transition at \emph{T}$_\texttt{O}$ = \emph{T}$_\texttt{N}$ = (201.5 $\pm$ 0.25) K with a temperature hysteresis of 1-2 K. (2) The observation of coexisting O-T phases over a finite temperature range near the transition [Fig.\ \ref{LattParam}a$'$] is a clear-cut evidence that the structural transition is first order, in addition to the discontinuous jump of the O distortion. (3) Based on the observation of remnant O features in the scattering (up to almost 450 K) we suggest that the O-T transition may be an order-disorder transition at \emph{T}$_\texttt{O}$. In this scenario, the continuous change in observed AFM domains as well as the small size of the average ordered AFM moments can be partially attributed to the existence of a minor T phase below \emph{T}$_\texttt{O}$. (4) The temperature dependence of sublattice magnetization [i.e., intensity of the AFM (103)$_\texttt{O}$ peak] exhibits a sudden jump at the transition and changes continuously below $T_\texttt{O}$ upon cooling or warming, however, the large difference between cooling and warming suggests that the AFM transition is a FO transition. (5) The temperature dependence of the AFM (103)$_\texttt{O}$ peak position strictly follows the evolution of the O \emph{a} lattice parameter providing conclusive evidence that the in-plane AFM propagation vector is along the long O \emph{a} axis.

\begin{acknowledgments}

Ames Laboratory is supported by the U.S. Department of Energy under Contract No. DE-AC02-07CH11358.

\end{acknowledgments}


\end{document}